\documentclass[]{spie}  

\usepackage{amsmath,amsfonts,amssymb}
\usepackage{graphicx}
\usepackage{media9}
\usepackage[colorlinks=true, allcolors=blue]{hyperref}

\title{Dynamics and interactions of particles in a thermophoretic trap}
\author{Benjamin Foster}
\author{Frankie Fung}
\author{Connor Fieweger}
\author{Mykhaylo Usatyuk}
\author{Anita Gaj}
\author{B.J. DeSalvo}
\author{Cheng Chin}
\affil{James Franck Institute, Enrico Fermi Institute, Department of Physics, University of Chicago, Chicago, IL 60637, USA}

\authorinfo{Further author information: (Send correspondence to C.C.)\\C.C.: E-mail: cchin@uchicago.edu, Telephone: 1 773 702 7192}

\begin{document}
\maketitle

\begin{abstract}
We investigate dynamics and interactions of particles levitated and trapped by the thermophoretic force in a vacuum cell. Our analysis is based on footage taken by orthogonal cameras that are able to capture the three dimensional trajectories of the particles. In contrast to spherical particles, which remain stationary at the center of the cell, here we report new qualitative features of the motion of particles with non-spherical geometry. Singly levitated particles exhibit steady spinning around their body axis and rotation around the symmetry axis of the cell. When two levitated particles approach each other, repulsive or attractive interactions between the particles are observed. Our levitation system offers a wonderful platform to study interaction between particles in a microgravity environment.
\end{abstract}

\keywords{Thermophoresis, levitation, trapping}

\section{INTRODUCTION}
\label{sec:intro}  

A wide variety of schemes have been developed in the past decades to levitate and trap objects. A prominent example is the use of magnetic levitation to move vehicles with greatly reduced friction. In the laboratory, levitation of atoms, ions, molecules, materials, and even biological bodies with electric, magnetic, optical, \cite{Ashkin71} or acoustic \cite{Trinh1985} fields has opened up a wide range of research fronts to investigate weak interactions of matter in an effective micro-gravity environment. However, most of the above methods have limited applications on objects with particular electromagnetic properties, eg., paramagnetic, polarizable, or conducting. Acoustic levitation applies to more general material in an open system, but the air also introduces significant friction that can quickly damp out dynamics of the levitated objects.

Recently, we have shown stable levitation and trapping of generic particles in a low pressure cell \cite{Fung2017}. By maintaining a large temperature gradient in the vertical direction, a variety of objects ranging from ice particles, glass bubbles, ceramic spheres, polyethylene particles, and even lint can be stably levitated and confined in the space between the two plates. The levitation and trapping force, called the thermophoretic force \cite{2002Zheng}, is a result of higher momentum transfer on the object from collisions with air molecules from the warmer side of the cell.

In our experiment, the levitation apparatus consists of a vacuum cell held near room temperature except for the top side of the cell supporting a bucket filled with liquid nitrogen at 77~K. Various particles are loaded into the cell before its evacuation to an air pressure of 1$\sim$10~Torr and are ejected into the trapping regime where stable levitation is observed. Singly levitated polyethylene spheres are stably trapped near the symmetry axis. A good agreement between the measured levitation height and theoretical models is reported \cite{Fung2017}.

In this paper, we describe the levitation of one or several non-spherical objects, which display far more interesting dynamics than do spherical particles. By imaging the particles from different directions, we fully reconstruct the three-dimensional trajectories of the levitated objects. We find that non-spherical objects tend to spin vertically and rotate around the symmetry axis of the cell. The spinning is consistent with the former observation on ice particles \cite{vanEimeren2012}; the rotation is an interesting finding, which we argue comes from the Magnus effect in thermophoresis. When more than one particle is levitated, we observe from their trajectories interaction between the particles separated by as far as a few mm. We attribute the interactions to the electrostatic charges on the levitated particles.

\section{THERMOPHORETIC LEVITATION AND TRAPPING}
\label{sec:Theory}

Thermophoresis refers to dynamics of an object in a fluid with an inhomogeneous temperature field \cite{2002Zheng}. This force originates from the momentum flow of the fluid in association with thermal conduction, and can be non-zero even in the absence of net relative motion between the fluid and the object and without an imbalance of thermodynamic pressure.

It is not difficult to picture an energy or a momentum flow without a net particle flow, which occurs in the thermal conduction regime; it is, however, puzzling to see a net mechanical force on an object without a pressure imbalance since pressure is usually associated with the force in hydrodynamics.  Thermophoretic force arises precisely when the system deviates from the hydrodynamic regime. More precisely, it emerges in the molecular flow regime when the mean-free-path $\lambda$ of the fluid elements is comparable or larger than the object \cite{2002Zheng}. Taking our system as an example, an object receives momentum from collisions with air molecules from all directions. While molecules from the hot side of the cell are more energetic than those from the cold side, the molecular density is also lower on the hot side. Thus the total pressure is balanced. A thermophoretic force comes about when the mean-free path $\lambda$ becomes significant: from the hot side of the cell with lower density and larger $\lambda$, the molecules that collide with the object originate from positions further away from the object than the molecules from the cold side.  The hot molecules thus carry more than enough momentum to compensate for the lower density and give the object a net momentum kick toward the cold side. In our system, thermophoretic force offers both the vertical levitation and radial confinement, see Figure ~1 (a).

Theoretically, the thermophoretic force for a spherical object with radius $r$ is given by \cite{2002Zheng}

\begin{eqnarray}
F_{th}=-\xi(\mbox{Kn})\frac{r^2\kappa}{\bar{v}}\nabla T,
\end{eqnarray}

\noindent where the minus sign indicates that the force points against the temperature gradient $\nabla T$, $\kappa$ is the thermal conductivity of the fluid, $\bar{v}=\sqrt{2k_BT/m}$ is the mean velocity of the gas molecule, $\xi$ is a dimensionless factor that depends on the Knudsen number $\mbox{Kn}\equiv \lambda/r$. In the continuum limit where $\mbox{Kn}\rightarrow 0$ the thermophoretic force vanishes $\xi\rightarrow 0$; in the molecular flow regime $\mbox{Kn}>>1$, the force reaches the maximum with $\xi=16\sqrt{\pi}/15\approx 1.8906$ \cite{Waldmann1959}.

\begin{figure} [ht]
\centering
   \includegraphics[height=6.3cm]{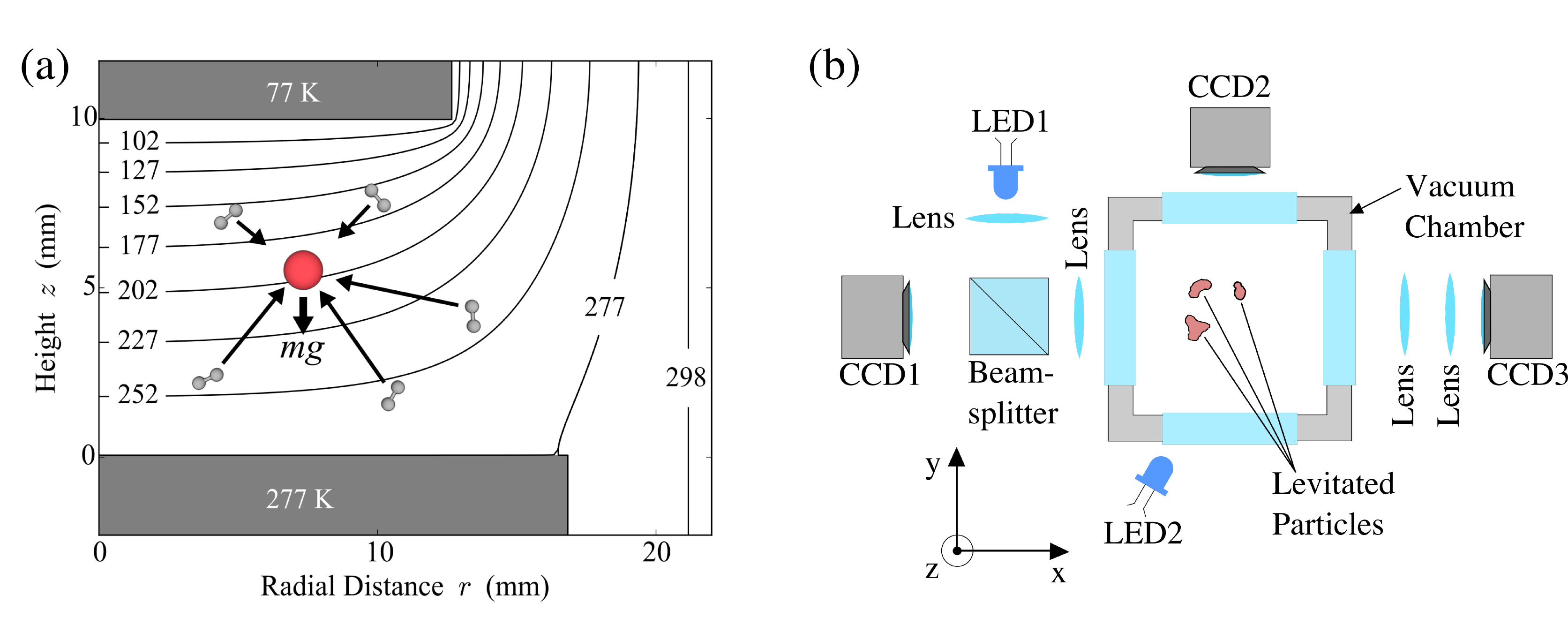}
   \caption[example]
{\label{Fig1}
Experimental thermal levitation scheme and detection setup. (a) The upper cold plate and the lower warm plate support an inhomogeneous temperature field, and particles are thermophoretically levitated between the two plates due to collisions with background molecules. (b) Our detection involves three cameras. CCDs (charge-coupled devices) 1 and 2 offer a large field of view to monitor the dynamics of the particles illuminated by LED 2 (light-emitting-diode). CCD 3 with collimated light from LED 1 performs absorption imaging with a close-up view to record the shapes and spin dynamics of the levitated particles.}
\end{figure}

Stable thermophoretic levitation in our system relies on the variation of $\xi$ in our cell. In the presence of a strong temperature gradient and properly tuned pressure, the gas near the hot bottom plate is deeper in the molecular flow regime and  $F_{th}$ is greater than the gravitational pull. The force, however, drops below gravity near the cold plate where gas is closer to the hydrodynamic regime. Stable levitation in the vertical direction thus occurs in the intermediate regime with $\mbox{Kn}\approx 1$. The crossover from molecular flow to hydrodynamic regime when the particles arise is the key to go beyond Earnshaw's theorem and stably levitate objects with a stationary temperature field \cite{Fung2017}.

Radial confinement is also required for three dimensional trapping of the levitated objects. In our system, this is provided by the geometry of the thermal cell that offers a positive curvature of the temperature distribution in the radial direction, see Figure 1 (a). Thus the thermophoretic force gives rise to a radial restoring force when the particle deviates from the symmetric axis of the system. A more quantitative analysis of the levitation and stability condition is given in Ref.~\cite{Fung2017}.

It should be noted that Eq.~(1) only applies to spheres. While levitation of generic objects with different shapes has been observed, an analytical expression for the levitation force on particles with general shape is unavailable \cite{2002Zheng}. It is thus the intention of this paper to identify new qualitative features in the dynamics of singly and multiply levitated particles with general, non-spherical shapes.

\section{EXPERIMENTAL APPARATUS}
\label{sec:apparatus}  

Our experimental apparatus has been described previously \cite{Fung2017}. At the center of the system are two circular metal plates spaced by 10.5~mm housed in a vacuum chamber. The top plate is cooled to 77~K by liquid nitrogen; the lower plate is left at room temperature. The temperature distribution in the cell is numerically calculated, see Figure 1 (a). We introduce particles between the plates by loading them on top of the lower plate. We then seal and evacuate the chamber to a pressure $P=1\sim$10~Torr. To levitate and trap the particles, we shake the cell until one or some of the particles launch from the bottom plate and reach the trapping zone.

We employ three cameras to record the dynamics of the particles, see Figure 1 (b). Two cameras in orthogonal directions offer a large field of view to track the dynamics of the particles, from which we can reconstruct the full three dimensional trajectories of particles. The third camera is placed beyond a microscope to visualize the shape and spinning of the particles. The frame rate for all cameras is 15 frames per second, which is sufficient to record the dynamics of the particles studied in this work.

\section{SINGLE-PARTICLE SPINNING AND ROTATION}
\label{sec:1particle}

Significantly different behaviors are observed for spherical and non-spherical particles levitated in our system. After launching, spherical particles quickly drift toward the equilibrium position, where they remain stable indefinitely. The heights of the particles in stationary state offer a quantitative test on Eq.~(1). An agreement with the theoretical calculations has been reported previously \cite{Fung2017}.

For non-spherical particles, the main target of this study, rich dynamics manifest even when they are singly levitated. Many of the levitated particles display a persistent spinning motion and rotation around the trap symmetry axis. More frequently than not, the spinning and rotation are along the $z-$axis, and can be seen under a large range of temperature and pressure in our experiment; see Figure~2 for an example. Such dynamics in steady state are interesting since the air friction at our pressure should damp out the motion in the time scale of a few ms. The motion thus suggests the particles are driven by a constant torque relative to both the body axis and the symmetry axis of the cell. In the following, we will separately discuss the origin of the spinning and rotation dynamics for non-spherical particles.

Spinning of levitated particles can be understood as a result of net torque on the levitated particles due to thermophoresis. The torque is expected on irregular objects since the phoretic force acts on the surface of the object, and the total force integrated over the surface does not necessarily go through the center of mass like gravity \cite{Rosner1989}. Such ``thermophoretic torque'' relative to the center-of-mass is generally expected to align the long axis of the particle with the temperature gradient \cite{Mackowski1990, Rosner1991}. Former experiments observed spinning of ice particles, which is interpreted as a result of the photophoretic force \cite{vanEimeren2012}. In our system, however, we showed that the photophoretic force is negligible \cite{Fung2017}.

The rotational motion of singly levitated particles is an unexpected feature in thermophoresis and is also found to be generic for irregular particles in our experiments. From three-dimensional (3D) trajectory reconstruction, the rotational motions are usually circular on the horizontal plane and very stable, see Figure 2. A wide range of rotation frequency $\omega/2\pi=0.1\sim 10$~Hz and amplitudes $0.1\sim 3$~mm are observed from different particles.

A steady circular motion relative to the trap center calls for a force that is both perpendicular to the radial restoring force pointing toward the trap center and the direction of the thermal gradient. Such force resembles the Magnus effect that lifts a spinning baseball. The conventional Magnus force is given by $F_{\mbox{Mag}}\propto \omega\times v$, where $v$ is the velocity of the object relative to the fluid and $\omega$ is its angular velocity. Here, we propose that a similar Magnus effect occurs in thermophoresis and can be responsible for the rotational motion of the spinning particles.

Microscopically, momentum transfer that leads to the thermophoretic force is not fundamentally different from that due to a relative velocity $v$ between the particle and the fluid. We may thus introduce an effective velocity $v\sim -\nabla T$ coming from the temperature gradient even in the case when the particle is not moving relative to the fluid. Thus a Magnus effect in thermophoresis for spinning particles can be modelled as  $F_{\mbox{Mag}}=-\alpha \omega\times\nabla T$, where $\alpha$ is a constant.

\begin{figure}[b]
\centering
 \includegraphics[width=.85\linewidth]{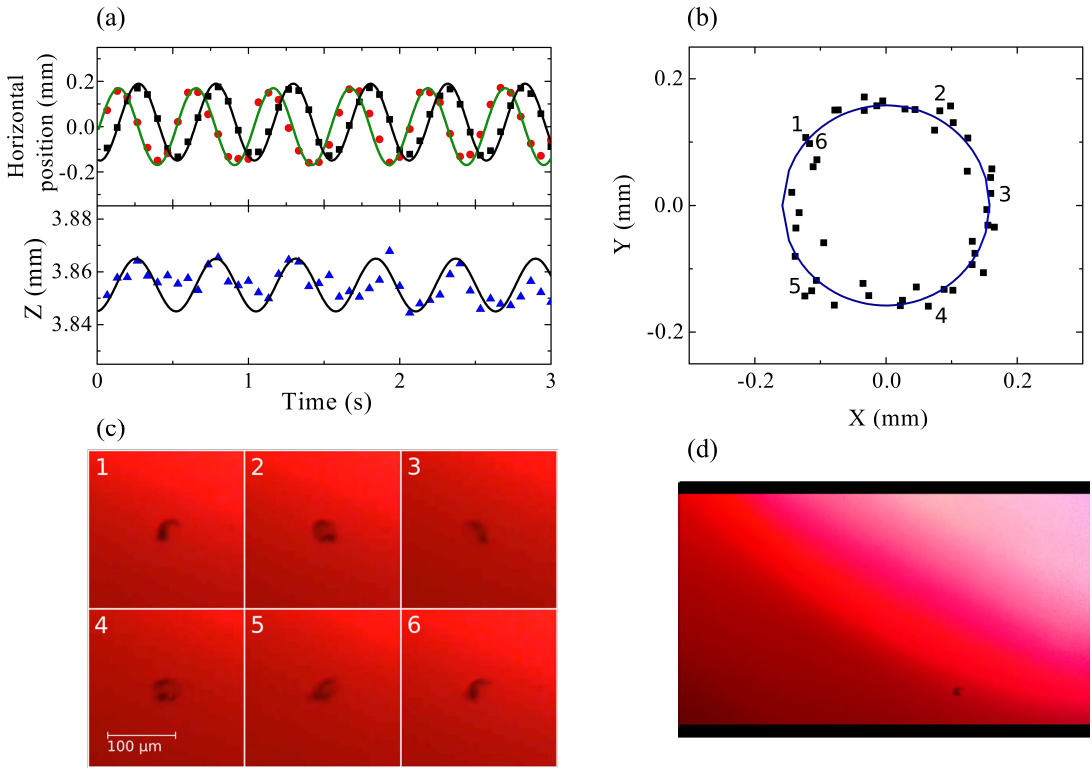}
 \caption[example]
{\label{Fig2}
Spinning and rotation of a singly levitated polyethylene particle. (a) Horizontal ($x$ and $y$) and vertical ($z$) positions of a single particle extracted from CCDs 1 and 2 (symbols) show oscillations as a function of time. From fitting the data in the range of $0\sim 3$~s (solid lines), we obtain the amplitude $A=0.15$~mm and frequency $\omega/2\pi=1.97$~Hz on the horizontal plane. An observed vertical motion about 10 times weaker may be due to a slight misalignment of the camera with respect to the plane of the particle's motion. (b) Reconstructed trajectory of the particle based on (a) (data: squares and line: fit) shows that the particle assumes a circular rotation about the symmetry axis of the system. Numbers $1\sim 6$ label the time sequence of the circulation. (c) Close-up view from CCD 3 shows the spinning of the same particle captured at times $1\sim6$. (d) Thumbnail for video in which dynamics of the particle based on the close up camera shows both spinning and rotation of the particle (full video in appendix).  The field of view is 1.75 mm by 1.1 mm.}
\end{figure}

\pagebreak
\section{TWO-PARTICLE INTERACTION}
\label{sec:2particles}

When more particles are levitated, interactions between them can be seen from the correlation of their dynamics. Here we discuss the interactions between 2 levitated polyethylene particles. To levitate exactly two particles, we carefully increase the strength that we shake the system until two particles enter the trapping zone. After both particles settle in the trap, we record their dynamics and reconstruct the 3D trajectories of both particles.

Figure 3 shows an example in which both levitated particles have non-spherical shape. The pressure of the vacuum cell is initially $>$7~Torr and later slowly reduces to 6~Torr in about 30~s. At first, the two particles are well separated by over 5~mm and their dynamics appear to be independent: particle $A$ rotates around the trap center at a characteristic frequency of $\omega_1/2\pi\approx 2$~Hz while particle $B$ at a much slower frequency $\omega_2/2\pi\approx0.05$~Hz. Amplitudes of their rotation are around $0.3$~mm. When the pressure reduces, the two particles approach each other and settle to a much smaller separation of $d=0.5$~mm.

At this small distance, the two particles influence each other much more, which can be clearly seen and analyzed from their trajectories, see Figure 3. The trajectory of the fast rotating particle (particle $A$) is modulated by the slow rotating one, and vice versa. The trajectories are well fit by an empirical form with two sinusoidal terms at $\omega_1$ and $\omega_2$. The fit further confirms that the modulation of one particle is synchronized with the rotation of the other particle. In particular, the slow modulation of particle $A$ is clearly in phase with particle $B$, which indicates an attraction between the two particles. In other similar experiments, we observe both attraction and repulsion between polyethylene, ceramic, and ice particles.

One possible candidate for the interactions is the electrostatic charges. Charges can be transferred to the particles by friction during the launching process. A rough estimation suggests that $100\sim 10,000$ electron charges ($e$) can be transferred to the particles. A future goal of the experiment is to verify that the force indeed scales as $1/d^2$, which can be checked by looking at the cross-talk between 2 particles at different interparticle separation $d$. If confirmed, an ensemble of many charged particles in levitation can potentially host a wealth of interesting many-body phenomena.
\begin{figure} [h]
 \centering
  \includegraphics[width=\linewidth]{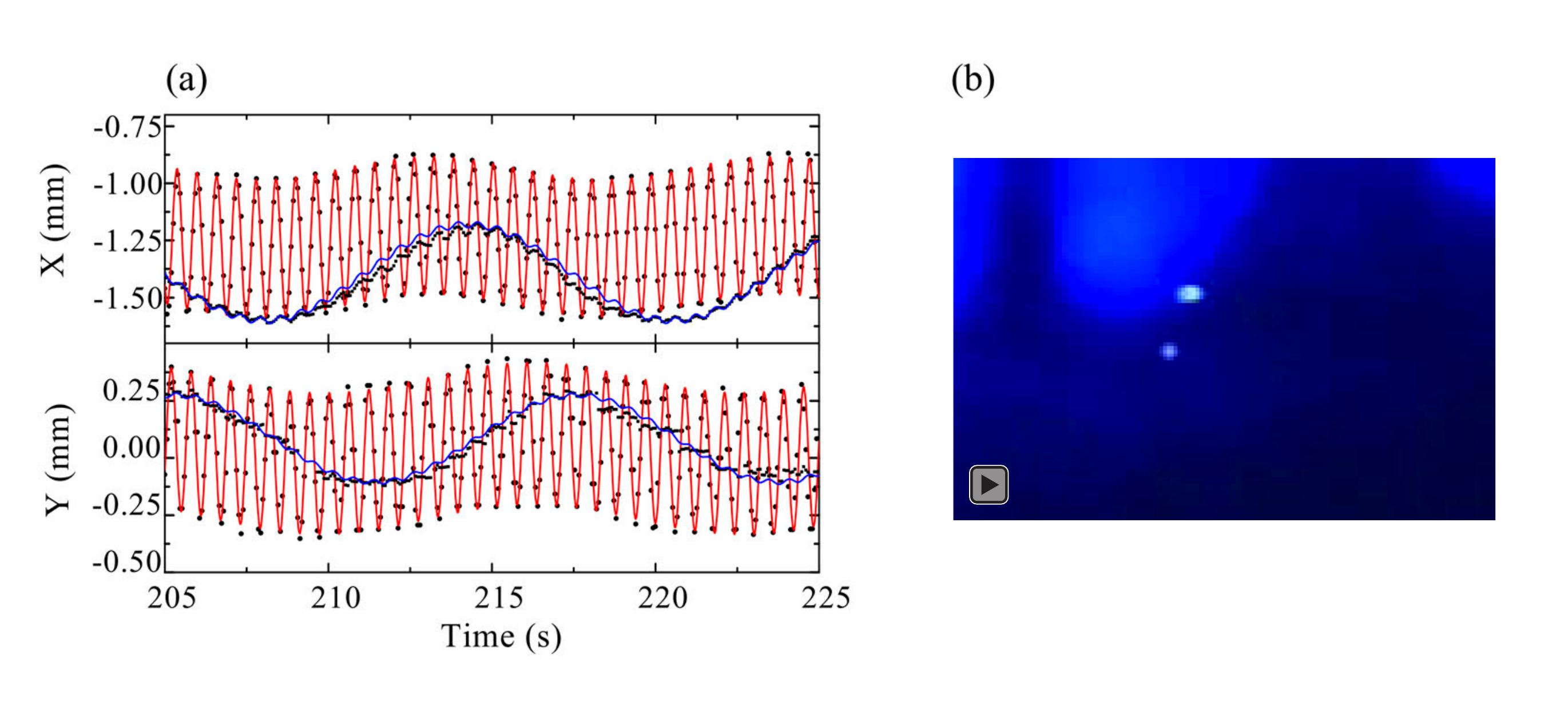}
 \caption[example]
{ \label{fig:example}
Dynamics of two levitated particles. (a) Horizontal ($x$ and $y$) dynamics of the particles extracted from cameras 1 and 2 (solid circles) shows oscillations at two frequencies $\omega_1$ and $\omega_2$. We fit each particle individually using the fit function $f(t)=a_1 \cos(\omega_1 t+\phi_1)+a_2 \cos(\omega_2 t+\phi_2)$ (solid lines). Particle $A$ predominately oscillates at frequency $\omega_1/2\pi=1.65$~Hz with amplitude $a_1=0.29$~mm; particle $B$ mainly oscillates at $\omega_2/2\pi=0.57$~Hz and $a_2=0.19$~mm. The small cross talk between the particles comes from the interaction of the particles. (b) Thumbnail for video demonstrating two particle interaction (full video in appendix).  The field of view is 6 mm by 3.2 mm}
\end{figure}

\section{Conclusion}
\label{sec:2particles}

In this paper we report a preliminary study of the dynamics of one and two polyethylene particles with irregular shape in a thermophoretic trap. Multiple cameras provide 3D reconstruction of their trajectories and a close-up view. A generic feature of singly levitated particles is their persistent spinning and rotation. The former can be understood as a thermophoretic torque on a non-spherical object and the latter is likely a new Magnus effect on a spinning particle in the presence of an inhomogeneous temperature field. When 2 particles are levitated, repulsive and attractive interactions can be seen from the cross-talk of their rotational motion. A potential origin of the interaction is the electrostatic charges on the particles. The dynamics we observe on the one- and two-body level suggests a wealth of intriguing physics in an ensemble of many levitated particles, an interesting topic for future research.

\acknowledgments 

We thank C. Parker and N. Kowalskii for the assistance in the early phase of the experiment. We acknowledge T. Witten for useful discussions. B.~J.~D. acknowledges support from the Grainger postdoc fellowship. A.~G. acknowledges support from the Kadanoff-Rice postdoc fellowship. B.~F acknowledges support from the James Frank Institute summer fellowship.  C.~F. acknowledges support from the Enrico Fermi Institute summer fellowship. This work was primarily supported by the University of Chicago Materials Research Science and Engineering Center, which is funded by National Science Foundation under award number DMR-1420709.


\bibliographystyle{spiebib} 

\pagebreak
\appendix
\section*{APPENDIX: Dynamics and Interactions of Levitated Particles}
\textbf{1. Figure 2(d) (Video length 10 seconds) }

\centering
\includemedia[
	label=fig2d,
	width=.6\linewidth,
	height=.2\paperheight,
	playbutton=plain,
	activate=pageopen,
	addresource=Figure2d.mp4,
	flashvars={source=Figure2d.mp4
		&autoplay=true
		&loop=false
	}
]{}{VPlayer.swf}	

\flushleft

\textbf{2. Figure 3(b) (Video length 21 seconds) }

\centering
\includemedia[
	label=fig3b,
	width=.6\linewidth,
	height=.2\paperheight,
	playbutton=plain,
	activate=pageopen,
	addresource=Fig3b.mp4,
	flashvars={source=Fig3b.mp4
		&autoplay=true
		&loop=false
	}
]{}{VPlayer.swf}	

\end{document}